\documentclass[aps,prd,showpacs,showkeys,twocolumn,10pt]{revtex4}

\usepackage{amssymb}
\usepackage{amsmath}
\usepackage[T1]{fontenc}
\usepackage{multirow}
\usepackage[breaklinks]{hyperref}

\begin{document}

\title{Magnetic properties of ground-state mesons}


\author{V.~\v{S}imonis}
 \email{vytautas.simonis@tfai.vu.lt}

\affiliation{
Vilnius University Institute of Theoretical Physics and
Astronomy,  Saul\.{e}tekio 3, LT-10222,\ Vilnius, Lithuania\\}

\date{\today}

\begin{abstract}
Starting with the bag model a method for the study of the magnetic
properties (magnetic moments, magnetic dipole transition widths) of
ground-state mesons is developed. We calculate the M1 transition moments and
use them subsequently to estimate the corresponding decay widths. These are
compared with experimental data, where available, and with the results
obtained in other approaches. Finally, we give the predictions for the
static magnetic moments of all ground-state vector mesons including those
containing heavy quarks. We have a good agreement with experimental data 
for the M1 decay rates of light as well as heavy mesons.
Therefore, we expect our predictions for the static magnetic properties (\textit{i.e.}, usual magnetic moments) to be of sufficiently high quality,
too.
\end{abstract}
\maketitle
\section{Introduction}
\label{sec_int}

The magnetic moments are among the fundamental properties of every hadron. 
They play an important role in the understanding of the hadronic structure. 
Theoretically the magnetic moment is related to the magnetic form factor of the hadron. For instance, it can be obtained by the extrapolation of the magnetic form factor $G_{M}(Q^{2})$ to zero momentum transfer. 
In the face of their importance, the magnetic moments of mesons have not
received much interest except for the $\rho ^{+}$ meson whose properties
were studied quite intensively using various approaches \cite
{S03,AOS09,HKLLWZ07,LMW08,GeA08,BS13,BM08,IKR99,HP99,HM98,BCJ02,CJ04,CGNSS95,J03,dF97,MS02,DEGM14,OKLMM15,LCD15}%
. The $K^{*}$ mesons have received less attention \cite
{AOS09,HKLLWZ07,LMW08,BS13,BM08,HP99,LCD15}, and we have found only three attempts 
\cite{LCD15,BS80,L03} to give some estimates for the magnetic moments of heavy
mesons. We think it is timely to pay the debt. Therefore, the study of the
magnetic moments of heavy mesons was the initial object of our
investigation. Because of the short lifetime the direct measurement of the
magnetic moments of vector mesons seems to be hardly possible. The indirect
estimate is possible \cite{SG14}, but it still suffers from the large
uncertainties. So we need some other observable to check the reliability of
our predictions. To this end we study the magnetic dipole (M1) transitions
of these mesons. In this case the experimental situation is better \cite
{PDG14}. In addition, there are also plenty of theoretical predictions
obtained using various approaches comprising the quark model and the vector
dominance model \cite{EK76,TK85,ST87,KX92,CDN93,J91,J96,IV95},
nonrelativistic QCD \cite{BeA05,BJV06}, the potential model \cite
{EGKLY78,EGKLY80,EQ94,SR80,BJ80,BJ82,HI82,O84,GOS84,ZSG91,FM93,GKLT95,LNR99,F99,BGS05,CS05,LS07,PPV10,DR11,DR12}%
, various relativistic or semirelativistic models \cite
{BGS05,GI85,G04,DS87,BDP92,BD94,JPM98,JPT99,JPS01,JPT02,JMPS10a,JMPS10b,ER93,A94,TZ94,CDN94,DFS00,CK01,EFG02,EFG03,RRS09,SPV14}%
, the bag model (including chiral extensions) \cite
{HDDT78,IDS82,WMM85,SM86,MS88,SM89,ZCT93,OH99}, the light front quark model \cite
{OX94,CGNSS95a,CJ99,C07}, models based on Bethe-Salpeter equation \cite
{L03,HS81,KRMMP00,LNR00}, QCD sum rules \cite
{RF79,K80,DN96,ZHY97,AIP94,ADIP96,S80,A84,BR85,AIP94a}, numerical
simulations of lattice QCD \cite{DER06,BH11,DeA12a,OKLMM15a}, chiral models 
\cite{CG92,AeA92,CCLLYY93,CCLLYY94,GR01,BEH03,HE05,CGZ15}, the statistical model \cite{BDS89}, the collective string-like model \cite{IK92}, the Nambu-Jona-Lasinio model \cite{BBHMR93,DCD14}, the constituent quark-meson model \cite{DDGNP98}, and the dispersion relation approach \cite{ADMNS07}. 

From a theoretical standpoint a static magnetic moment of a single quark is
the long wavelength limit of the M1 transition (spin-flip) moment of this
quark. Hence, these magnetic properties of hadrons are closely related, and,
if we succeeded in predicting M1 decay rates, we would get some confidence
that the predictions for magnetic moments were also reliable. We are going
to implement this plan with the help of the modified bag model \cite{BS04}
which was used earlier to calculate masses of light and heavy hadrons \cite
{BS04,BS09}, magnetic moments \cite{BS13a}, and M1 decay widths \cite{BS13b}
of heavy baryons.

The remainder of the paper is as follows. In sect.~\ref{sec_bag} the short description of our version of the bag model is given, and the formalism we use to treat the magnetic properties of the hadrons is presented. In sect.~\ref
{sec_cal} the predictions for the M1 transitions moments and partial decay
widths are given. They are compared with the results obtained in other
approaches and with experimental data. Our predictions for the magnetic
moments of ground-state vector mesons are presented in sect.~\ref{sec_rad}.
The last section serves for the summary and discussion.

\section{Bag model and magnetic observables}
\label{sec_bag}

The MIT bag model in the static cavity approximation \cite{DJJK75} is a
simple intuitive approach to hadron structure (see also the excellent review 
\cite{T84}). We use the modified version \cite{BS04} designed to reconcile
the initial ultrarelativistic model with the heavy quark physics. Here we
review the main features of this model (for details we refer to ref.~\cite
{BS04}).

It is assumed that quarks are confined in the sphere of radius $R$, within
which they obey the free Dirac equation. The four-component wave function of
the quark in the $s$-mode is given by 
\begin{equation}
\Psi _{\ m}^{1/2}(r)=\frac{1}{\sqrt{4\pi }}\left( 
\begin{array}{c}
G(r) \\ 
-i(\mathbf{\sigma }\cdot \mathbf{\hat{r})}F(r)
\end{array}
\right) \Phi _{\ m}^{1/2}\ ,  \label{bag 001}
\end{equation}
where $\Phi _{\ m}^{1/2}$ is two-component spinor, $\mathbf{\sigma }$ are
usual Pauli matrices, and $\mathbf{\hat{r}}$ is unit radius-vector.
Solutions of the free Dirac equation in the spherical coordinate system are
simple Bessel functions, so that 
\begin{equation}
G(r)=Nj_{0}(pr)\ ,  \label{bag 002}
\end{equation}

\begin{equation}
F(r)=-N\sqrt{\frac{\varepsilon -m}{\varepsilon +m}}\ j_{1}(pr)\ ,
\label{bag 003}
\end{equation}
where $\varepsilon ^{2}=p^{2}+m^{2},$ and 
\begin{equation}
j_{0}(x)=\frac{\sin x}{x}\ ,\ j_{1}(x)=\frac{\sin x}{x^{2}}-\frac{\cos x}{x}%
\ .  \label{bag 004}
\end{equation}

The normalization factor $N$ is 
\begin{equation}
N=\frac{p^{2}}{\sin (pR)\sqrt{2\varepsilon \left( \varepsilon R-1\right) +m}}%
\ .  \label{bag 005}
\end{equation}

The eigenenergy of the quark is determined by the matching condition at the
bag boundary

\begin{equation}
G(R)=-F(R)\ ,  \label{bag 006}
\end{equation}
from which using eqs.~(\ref{bag 002})--(\ref{bag 004}) one obtains

\begin{equation}
\tan (pR)=\frac{pR}{1-mR-\varepsilon R}\ .  \label{bag 007}
\end{equation}

The energy of the bag associated with a particular hadron is given by

\begin{equation}
E=\frac{4\pi }{3}BR^{3}+\sum\limits_{i}\varepsilon _{i}+E_{\mathrm{int}}\,,
\label{bag 008}
\end{equation}
where $R$ denotes the bag radius, and $B$ is the bag constant. The entries
on the right-hand side of this expression are the bag volume energy, the sum
of single-particle eigenenergies, and the quark-quark interaction energy due
to one-gluon-exchange. Minimization of the energy (\ref{bag 008}) determines
the bag radius $R_{\mathrm{H}}$ of particular hadron. $E_{\mathrm{int}}$
represents the interaction energy of quarks in the Abelian approximation to
QCD and is comprised of color-electric and color-magnetic parts as described
in ref. \cite{BS04} in detail. Equation~(\ref{bag 008}) differs from the usual
MIT bag energy in that we have omitted the Coulomb-type self-energy and
Casimir energy terms. The interaction energy $E_{\mathrm{int}}$ depends on
the running coupling constant, for which we employ the parametrization
proposed in ref. \cite{DJ80}

\begin{equation}
\alpha _{\mathrm{c}}(R)=\frac{2\pi }{9\,\ln (A+R_{0}/R)}\,\,.
\label{bag 009}
\end{equation}

In eq.~(\ref{bag 009}) $R_{0}$ is the scale parameter analogous to the QCD
constant $\Lambda $. The parameter $A$ helps to avoid divergences when $%
R_{0}\rightarrow R$. Other differences from the original MIT bag model are
the effective (running) quark mass 
\begin{equation}
m_{f}(R)=\widetilde{m}_{f}+\alpha _{\mathrm{c}}(R)\cdot \delta _{f}\,,
\label{bag 010}
\end{equation}
and the corrections for the center-of-mass motion (CMM). Parameters $%
\widetilde{m}_{f}$ and $\delta _{f}$ are necessary to define the mass
functions $m_{f}(R)$ for each quark flavor.

The CMM corrected bag energy $M$ is identified with the mass of the hadron. It
is related to the uncorrected energy $E$ by

\begin{equation}
M^{2}=E^{2}-P^{2},  \label{bag 011}
\end{equation}
where 
\begin{equation}
P^{2}=\gamma \sum\limits_{i}p_{i}^{2}  \label{bag 012}
\end{equation}
is the effective momentum square, $\gamma $ is the parameter governing the
CMM prescription, $p_{i}=\sqrt{\varepsilon _{i}^{2}-m_{i}^{2}}$ represent
momenta of individual quarks, and $m_{i}$ is the effective quark mass given by
eq.~(\ref{bag 010}).

To fix the model parameters $B$, $\gamma $, $A$ and $R_{0}$ the
experimentally observed masses of the light hadrons $N$, $\Delta $, $\pi $
and $\rho $ were chosen. To fix the mass function parameters $\tilde{m}_{s}$%
, $\delta _{s}$, $\tilde{m}_{c}$, $\delta _{c}$, $\tilde{m}_{b}$, $\delta
_{b}$ we have employed the masses of vector mesons ($\phi ,$ $J/\psi ,$ 
$\Upsilon$) and the mass values of the corresponding lightest 
baryons ($\Lambda ,$ $%
\Lambda _{c},$ $\Lambda _{b}$). The lightest (up and down) quarks are assumed
to be massless. The fitted numerical values are $B=7.301\times 10^{-4}~%
\mathrm{GeV}^{4}$, $\gamma =1.785$, $A=0.7719$, $R_{0}=3.876~\mathrm{GeV}%
^{-1}$, $\widetilde{m}_{s}=0.2173~\mathrm{GeV}$, $\delta _{s}=0.1088$~$%
\mathrm{GeV}$, $\widetilde{m}_{c}=1.456~\mathrm{GeV}$, $\delta _{c}=0.1003~%
\mathrm{GeV}$, $\widetilde{m}_{b}=4.746~\mathrm{GeV}$, and $\delta
_{b}=0.0880~\mathrm{GeV}$.

The magnetic moment of a hadron is obtained from the usual definition 
\begin{equation}
\mathbf{\mu }=\frac{1}{2}\int \mathrm{d}^{3}x\left[ \mathbf{r}\times \mathbf{%
j}_{em}\right] ,  \label{bag 013}
\end{equation}
where $\mathbf{j}_{em}$ is the Dirac electromagnetic current (see ref.~\cite
{T84}). After some simple algebra one obtains 
\begin{equation}
\mathbf{\mu } =\sum \widetilde{\mu }_{i}\left\langle h\uparrow \left| e_{i}\mathbf{\sigma }
_{i}\right| h\uparrow \right\rangle ,  \label{bag 014}
\end{equation}
where $e_{i}$ is the charge of the corresponding quark and $\widetilde{\mu }%
_{i}$ is it's reduced (charge-independent) magnetic moment 
\begin{equation}
\widetilde{\mu }_{i}=\int r^{2}\mathrm{d}r\frac{2r}{3}G_{i}(r)F_{i}(r).
\label{bag 015}
\end{equation}

Upon evaluating the above matrix element for the reduced magnetic moment of
a quark confined in the bag of radius $R_{\mathrm{H}}$ we obtain 
\begin{equation}
\widetilde{\mu }_{i}=\frac{1}{6}\frac{2R_{\mathrm{H}}(2\varepsilon
_{i}+m_{i})-3}{2\varepsilon _{i}(\varepsilon _{i}R_{\mathrm{H}}-1)+m_{i}}\ .
\label{bag 016}
\end{equation}

The calculation of spin-flavor matrix elements is somewhat lengthy but
straightforward task. Using the technique described in ref.~\cite{C79} for
the vector meson made of quark $q_{a}$ and antiquark $\overline{q}_{b}$ we
obtain 
\begin{equation}
\mu =e_{a}\widetilde{\mu }_{a}+\overline{e}_{b}\widetilde{\mu }_{b},
\label{bag 017}
\end{equation}
where $e_{a}$ ($\overline{e}_{b}$) is the charge of the corresponding quark
(antiquark). Note that $e_{a}\widetilde{\mu }_{a}$ ($\overline{e}_{b}%
\widetilde{\mu }_{b}$) in the last equation represents the magnetic moment
of the individual quark (antiquark).

For the M1 decay widths of vector mesons we use the expression derived in 
the ref.~\cite
{HDDT78}, though written in a slightly different fashion 
\begin{equation}
\Gamma _{V\rightarrow PS}=\frac{4\alpha k^{3}}{3}(\mu _{V\rightarrow
PS})^{2}\,,  \label{bag 020}
\end{equation}
where $k=(M_{V}^{2}-M_{PS}^{2})/2M_{V}$ is the photon momentum in the rest
frame of decaying a particle, $\alpha \approx \frac{1}{137}$ is the fine
structure constant, and the transition moments connecting vector and
pseudoscalar mesons are 
\begin{equation}
\mu _{V\rightarrow PS}=e_{a}\widetilde{\mu }_{a}^{TR}-\overline{e}_{b}%
\widetilde{\mu }_{b}^{TR}.  \label{bag 018}
\end{equation}

The reduced transition moment $\widetilde{\mu }_{i}^{TR}$ of an individual
quark depends on the momentum of the emitted photon $k$ and is given by 
\begin{equation}
\widetilde{\mu }_{i}^{TR}(k)=\int r^{2}\mathrm{d}%
rj_{1}(kr)[G_{1i}(r)F_{2i}(r)+G_{2i}(r)F_{1i}(r)].  \label{bag 019}
\end{equation}

Indices $1$ and $2$ denote the initial and final particle. In the limit $%
k\rightarrow 0$, with $R_{1}=R_{2}$, $G_{1}(r)=G_{2}(r)$, and $%
F_{1}(r)=F_{2}(r),$ eq.~(\ref{bag 019}) reproduces the expression (\ref{bag
015}) for the static magnetic moment.

For the magnetic dipole decay of the pseudoscalar meson instead of eq.~(\ref
{bag 020}) we have the expression 
\begin{equation}
\Gamma _{PS\rightarrow V}=4\alpha k^{3}(\mu _{PS\rightarrow V})^{2}\,,
\label{bag 021}
\end{equation}
with $k=(M_{PS}^{2}-M_{V}^{2})/2M_{PS}$. In this case the transition moment $%
\mu _{PS\rightarrow V}$ is given by eqs.~(\ref{bag 018}) and (\ref{bag 019}%
), but with the indices $1$ and $2$ interchanged.

In evaluating the integral in (\ref{bag 019}) it is necessary to choose the
value of the upper limit of the integral. It is not a trivial procedure
because, in general, the bag radii of the particles under transition are
different. The choice in ref.~\cite{HDDT78} was to take $%
R=(R_{1}+R_{2})/2$. In the present treatment we prefer to use the smaller of the
two in order to take approximately into account the overlap of bags.

So far we have ignored any dynamical recoil effects and CMM corrections to
the magnetic (or magnetic transition) moments and decay rates.
Unfortunately, because of the fundamental difficulties, there is no rigorous
formalism available to account for the CMM and recoil corrections in the
relativistic constituent particle models such as the bag model. In general, these corrections
are model dependent and plausibly should be treated simultaneously.
Strictly speaking, any treatment of these corrections is nothing more than a
possible approximate method. In the case of light hadrons the CMM correction
is expected to play the main role. It is known that in the bag model
calculation of nucleon magnetic moment the recoil correction is relatively
small ($\leq 10\%$) and negative \cite{BG83,G83}. In the calculations of the
form factors and magnetic moments of octet baryons using cloudy bag model 
\cite{YTKK89} the typical recoil corrections also do not exceed $10\%$. If
one decides to ignore the recoil, for the CMM corrections of the static
magnetic moments one can use the simple approximate expression proposed by
Halprin and Kerman \cite{HK82} 
\begin{equation}
\mu =\frac{E}{M}\mu ^{0}\,,  \label{cal 023}
\end{equation}
where $\mu $ is the CMM corrected magnetic moment, $\mu ^{0}$ is the usual
uncorrected one, $M$ is the mass of the hadron, and $E$ represents the
uncorrected bag energy. However, for the transition moments this approach can not be applied literally because
the ratios $E_{i}/M_{i}$ for the initial and final mesons differ (sometimes
vastly). It could be interesting to test the model using more sophisticated
CMM correction schemes, such us suggested by Peierls and Yoccoz \cite{PY57},
for example. But such treatment is rather complicated and for the moment is
outside the scope of present investigation. Instead, we resort to the
rescaling procedure adopted in some bag model calculations \cite{IDS82,MS88}%
. It was also suggested \cite{YTKK89} that for the light baryon octet the
effect of the CMM correction on the magnetic form factors can be simulated
reasonably well by multiplying them with an overall constant. It is evident
that we can not use a single scale factor for all hadrons. In what follows we
assume that the integrated effect of the CMM, recoil, and possibly other
corrections on the magnetic observables (such as magnetic moments or M1
transition moments) can be simulated by a simple rescaling of the quark level
quantities. In practice as minimal prescription we have used the simplest 
possible ansatz 
\begin{equation}
\mu _{L}=C_{L}\,\mu _{L}^{0}\,,\quad \mu _{H}=C_{H}\,\mu _{H}^{0}\,,
\label{cal 024}
\end{equation}
where $\mu _{L}^{0}$ and $\mu _{H}^{0}$ represent the initial bag model
quantities (magnetic or M1 transition moments) for the light ($u,$ $d$, or $s
$) and heavy ($c$ or $b$) quarks, respectively. In short, our prescription
is to calculate various magnetic observables using usual expressions (such
as eqs.~(\ref{bag 017})--(\ref{bag 021})), but with magnetic observables
of individual quarks replaced by the corrected ones. The scale factor $C_{L}$
has been chosen to reproduce the experimental value of the magnetic moment
of the proton. Note that the same prescription was used in Ref.~\cite{IDS82}. In
our model we have $C_{L}=1.43$. The scale factor for the heavy quarks $C_{H}$
is essentially a new parameter. It was adjusted to reproduce the experimental
values of the transition moments $\mu _{D^{*+}\rightarrow D^{+}}$ and $\mu
_{J/\psi \rightarrow \eta _{c}}$ simultaneously. We have found that
acceptable values of $C_{H}$ span the range $0.82-0.92$ and have chosen as
optimal the central value $C_{H}=0.87$. As we see, the scale factor for the
heavy quarks $C_{H}$ is appreciably smaller than $C_{L}$. This could mean
that in the case of heavy quarks the CMM correction tends to vanish and the
recoil plausibly gains an advantage. Actually we do not know if we can use
for the strange quarks the same scale factor that was adjusted for the
lightest ($u$ and $d$) quarks and for the bottom quarks the same scale
factor that was adjusted for the charmed quarks. At present it seems to be a
reliable choice. Either way, the magnetic moment of the bottom quark is
relatively small, and therefore tiny changes in it do not affect other
predictions greatly.

\section{Magnetic dipole transitions}
\label{sec_cal}

First, for convenience, we present in table~\ref{cal3.01} the
quark-antiquark structure of $s$-state mesons. Conjugate particles can be
obtained from those listed in the table by the substitution $%
q_{a}\rightarrow \overline{q}_{a},\overline{q}_{b}\rightarrow q_{b}$.

\begin{table}[tbp] \centering 
\caption{Spin-flavor content of ground-state mesons.} 
\label{cal3.01}
\begin{tabular}{ccc}
\hline\noalign{\smallskip}
Flavor content & $J=0$ & $J=1$ \\ 
\noalign{\smallskip}\hline\noalign{\smallskip}
$-u\overline{d}$ & $\pi ^{+}$ & $\rho ^{+}$ \\ 
$(u\overline{u}-d\overline{d})/\sqrt{2}$ & $\pi ^{0}$ & $%
\rho ^{0}$ \\ 
$(u\overline{u}+d\overline{d})/\sqrt{2}$ & $\eta _{l}$ & 
$\omega _{l}$ \\ 
$u\overline{s}$ & $K^{+}$ & $K^{*+}$ \\ 
$d\overline{s}$ & $K^{0}$ & $K^{*0}$ \\ 
$-s\overline{s}$ & $\eta _{s}$ & $\phi _{s}$ \\ 
$c\overline{d}$ & $D^{+}$ & $D^{*+}$ \\ 
$c\overline{u}$ & $D^{0}$ & $D^{*0}$ \\ 
$c\overline{s}$ & $D_{s}^{+}$ & $D_{s}^{*+}$ \\ 
$c\overline{c}$ & $\eta _{c}$ & $J/\psi $ \\ 
$u\overline{b}$ & $B^{+}$ & $B^{*+}$ \\ 
$d\overline{b}$ & $B^{0}$ & $B^{*0}$ \\  
$s\overline{b}$ & $B_{s}^{0}$ & $B_{s}^{*0}$ \\ 
$c\overline{b}$ & $B_{c}^{+}$ & $B_{c}^{*+}$ \\  
$b\overline{b}$ & $\eta _{b}$ & $\Upsilon$ \\ 
\noalign{\smallskip}\hline
\end{tabular}
\end{table} 

We assume the physical states of pseudoscalar ($\eta ,\eta ^{\prime }$) and
vector ($\omega ^{0},\phi $) mesons to be the mixtures of the ($\eta
_{l},\eta _{s}$) and ($\omega _{l},\phi _{s}$) states

\begin{equation}
\left. 
\begin{array}{c}
\eta =-\eta _{l}\,\sin \alpha _{P}+\eta _{s}\,\cos \alpha _{P}\,, \\[1ex] 
\eta ^{\prime }=\eta _{l}\,\cos \alpha _{P}+\eta _{s}\,\sin \alpha _{P}\,,
\end{array}
\right.  \label{cal 001}
\end{equation}

\begin{equation}
\left. 
\begin{array}{c}
\omega ^{0}=\omega _{l}\,\cos \alpha _{V}+\phi _{s}\,\sin \alpha _{V}\,, \\ 
\phi =-\omega _{l}\,\sin \alpha _{V}+\phi _{s}\,\cos \alpha _{V}\,.
\end{array}
\right.  \label{cal 002}
\end{equation}

Definitions of the mixed states and phase systems of the wave functions used
by various authors may differ. Ours are the same as in ref.~\cite{O81}.

Strictly speaking, the physical states ($\eta ,\eta ^{\prime }$) and ($%
\omega ^{0},\phi $) can also contain the admixtures of radial excitations,
heavier quarkonium states, and glue \cite{CL79,R83,F00}. Because the 
state mixing problem is not our main subject of interest, we confine ourselves with the simplified picture given by eqs.~(\ref{cal 001}) and (\ref{cal 002}). If not explicitly stated otherwise, we will
determine the mixing angle $\alpha _{V}$ empirically from the fit to the
data. In our model setting $\alpha _{V}=-4.1^{\circ }$ we have $\mu _{\phi
\rightarrow \pi ^{0}}=0.124~\mu _{N}$ consistent with the experimental value 
$0.125\pm 0.004~\mu _{N}$. Here and in what follows $\mu _{N}=e/(2M_{P})$ is
the nuclear magneton, and $M_{P}$ is the mass of the proton. For the
pseudoscalars we take the widely accepted \cite{GI85,O81} perfect mixing angle $%
\alpha _{P}=-45^{\circ }$. This choice is useful if we want to compare our
predictions with the results obtained using other approaches in which the
same perfect mixing angle was used.

With these preliminaries by using eqs.~(\ref{bag 018}), (\ref{cal 001}), and (%
\ref{cal 002}) we can write down the
detailed expressions for the transition moments, i.e., 
\begin{equation}
\mu _{\rho ^{+}\rightarrow \pi ^{+}}(k)=\mu _{\rho ^{0}\rightarrow \pi
^{0}}(k)=\frac{1}{3}\widetilde{\mu }_{l}^{TR}(k)\,,  \label{cal 003}
\end{equation}
\begin{equation}
\mu _{K^{*+}\rightarrow K^{+}}(k)=\frac{2}{3}\widetilde{\mu }_{l}^{TR}(k)-%
\frac{1}{3}\widetilde{\mu }_{s}^{TR}(k)\,,  \label{cal 004}
\end{equation}
\begin{equation}
\mu _{K^{*0}\rightarrow K^{0}}(k)=-\frac{1}{3}\widetilde{\mu }_{l}^{TR}(k)-%
\frac{1}{3}\widetilde{\mu }_{s}^{TR}(k)\,,  \label{cal 005}
\end{equation}
\begin{equation}
\mu _{\omega ^{0}\rightarrow \pi ^{0}}(k)=\widetilde{\mu }_{l}^{TR}(k)\,\cos
\alpha _{V}\,,  \label{cal 006}
\end{equation}
\begin{equation}
\mu _{\phi \rightarrow \pi ^{0}}(k)=-\widetilde{\mu }_{l}^{TR}(k)\,\sin
\alpha _{V}\,,  \label{cal 007}
\end{equation}
\begin{equation}
\mu _{\rho ^{0}\rightarrow \eta }(k)=-\widetilde{\mu }_{l}^{TR}(k)\,\sin
\alpha _{P}\,,  \label{cal 008}
\end{equation}
\begin{equation}
\mu _{\eta ^{\prime }\rightarrow \rho ^{0}}(k)=\widetilde{\mu }%
_{l}^{TR}(k)\,\cos \alpha _{P}\,,  \label{cal 009}
\end{equation}
\begin{eqnarray}
\mu _{\omega ^{0}\rightarrow \eta }(k) &=&-\frac{1}{3}\widetilde{\mu }%
_{l}^{TR}(k)\,\cos \alpha _{V}\,\sin \alpha _{P}  \nonumber \\
&&-\frac{2}{3}\widetilde{\mu }_{s}^{TR}(k)\,\sin \alpha _{V}\,\cos \alpha
_{P}\,,  \label{cal 010}
\end{eqnarray}
\begin{eqnarray}
\mu _{\eta ^{\prime }\rightarrow \omega ^{0}}(k) &=&\frac{1}{3}\widetilde{%
\mu }_{l}^{TR}(k)\,\cos \alpha _{V}\,\cos \alpha _{P}  \nonumber \\
&&-\frac{2}{3}\widetilde{\mu }_{s}^{TR}(k)\,\sin \alpha _{V}\,\sin \alpha
_{P}\,,  \label{cal 011}
\end{eqnarray}
\begin{eqnarray}
\mu _{\phi \rightarrow \eta ^{\prime }}(k) &=&-\frac{1}{3}\widetilde{\mu }%
_{l}^{TR}(k)\,\sin \alpha _{V}\,\cos \alpha _{P}  \nonumber \\
&&-\frac{2}{3}\widetilde{\mu }_{s}^{TR}(k)\,\cos \alpha _{V}\,\sin \alpha
_{P}\,,  \label{cal 012}
\end{eqnarray}
\begin{eqnarray}
\mu _{\phi \rightarrow \eta }(k) &=&\frac{1}{3}\widetilde{\mu }%
_{l}^{TR}(k)\,\sin \alpha _{V}\,\sin \alpha _{P}  \nonumber \\
&&-\frac{2}{3}\widetilde{\mu }_{s}^{TR}(k)\,\cos \alpha _{V}\,\cos \alpha
_{P}\,,  \label{cal 013}
\end{eqnarray}
\begin{equation}
\mu _{D^{*+}\rightarrow D^{+}}(k)=-\frac{1}{3}\widetilde{\mu }_{l}^{TR}(k)+%
\frac{2}{3}\widetilde{\mu }_{c}^{TR}(k)\,,  \label{cal 014}
\end{equation}
\begin{equation}
\mu _{D^{*0}\rightarrow D^{0}}(k)=\frac{2}{3}\widetilde{\mu }_{l}^{TR}(k)+%
\frac{2}{3}\widetilde{\mu }_{c}^{TR}(k)\,,  \label{cal 015}
\end{equation}
\begin{equation}
\mu _{D_{s}^{*+}\rightarrow D_{s}^{+}}(k)=-\frac{1}{3}\widetilde{\mu }%
_{s}^{TR}(k)+\frac{2}{3}\widetilde{\mu }_{c}^{TR}(k)\,,  \label{cal 016}
\end{equation}
\begin{equation}
\mu _{J/\psi \rightarrow \eta _{c}}(k)=\frac{4}{3}\widetilde{\mu }%
_{c}^{TR}(k)\,,  \label{cal 017}
\end{equation}
\begin{equation}
\mu _{B^{*+}\rightarrow B^{+}}(k)=\frac{2}{3}\widetilde{\mu }_{l}^{TR}(k)-%
\frac{1}{3}\widetilde{\mu }_{b}^{TR}(k)\,,  \label{cal 018}
\end{equation}
\begin{equation}
\mu _{B^{*0}\rightarrow B^{0}}(k)=-\frac{1}{3}\widetilde{\mu }_{l}^{TR}(k)-%
\frac{1}{3}\widetilde{\mu }_{b}^{TR}(k)\,,  \label{cal 019}
\end{equation}
\begin{equation}
\mu _{B_{s}^{*0}\rightarrow B_{s}^{0}}(k)=-\frac{1}{3}\widetilde{\mu }%
_{s}^{TR}(k)-\frac{1}{3}\widetilde{\mu }_{b}^{TR}(k)\,,  \label{cal 020}
\end{equation}
\begin{equation}
\mu _{B_{c}^{*+}\rightarrow B_{c}^{+}}(k)=\frac{2}{3}\widetilde{\mu }%
_{c}^{TR}(k)-\frac{1}{3}\widetilde{\mu }_{b}^{TR}(k)\,,  \label{cal 021}
\end{equation}
\begin{equation}
\mu _{\Upsilon
\rightarrow \eta _{b}}(k)=-\frac{2}{3}\widetilde{\mu }_{b}^{TR}(k)\,.
\label{cal 022}
\end{equation}
$\widetilde{\mu }_{l}^{TR}$ in the expressions above denote the reduced
transition moments of the lightest (up or down) quarks. Note that isospin symmetry implies $%
\widetilde{\mu }_{l}^{TR}=\widetilde{\mu }_{u}^{TR}=\widetilde{\mu }%
_{d}^{TR} $.

\begin{table*}[tbp] \centering%
\caption{Transition moments (in nuclear
magnetons) of ground-state vector mesons.} 
\label{cal3.02}
\begin{tabular}{lcrrrrccr}
\hline\noalign{\smallskip}
Transition & Experiment$^{a}$ & NR & LWL & Our & SRPM & RPM$^{a}$ & RPM$^{a}$
& SM \\ 
&  &  &  &  & \cite{GI85} & \cite{JPT02} & \cite{CK01} & \cite{BDS89} \\ 
\noalign{\smallskip}\hline\noalign{\smallskip}
$\rho ^{+}\rightarrow \pi ^{+}$ & $0.68\pm 0.04$ & $0.93$ & $0.82$ & $0.68$ & $0.69$ & $0.68$ & $0.62$ & $0.69$ \\ 
$\rho ^{0}\rightarrow \pi ^{0}$ & $0.78\pm 0.05$ & $0.93$ & $0.82$ & $0.68$ & $0.69$ & $0.68$ & $0.61$ & $0.69$ \\ 
$\omega ^{0}\rightarrow \pi ^{0}$ & $2.15\pm 0.04$ & $2.79$ & $2.47$ & $2.01$ & $2.07$ & $2.02$ & $1.82$ & $2.07$ \\ 
$\omega ^{0}\rightarrow \eta $ & $0.42\pm 0.02$ & $0.70$ & $0.64$ & $0.60$ & $0.50$ & $0.47$ & $0.49$ & $0.50$ \\ 
$\rho ^{0}\rightarrow \eta $ & $1.49\pm 0.05$ & $1.97$ & $1.75$ & $1.66$ & $1.53$ & $1.67$ & $1.66$ & $1.50$ \\ 
$K^{*+}\rightarrow K^{+}$ & $0.78\pm 0.04$ & $1.25$ & $0.97$ & $0.85$ & $0.91$ & $0.84$ & $0.80$ & $\cdot \cdot \cdot $ \\ 
$K^{*0}\rightarrow K^{0}$ & $1.19\pm 0.05$ & $-1.54$ & $-1.35$ & $-1.20$ & $-1.20$ & $1.17$ & $1.23$ & $\cdot \cdot \cdot $ \\ 
$\phi \rightarrow \pi ^{0}$ & $0.125\pm 0.004$ & $0.125$ & $0.18$ & $0.124$ & $0.06$ & 
$0.12$ & $0.092$ & $0.45$ \\ 
$\phi \rightarrow \eta $ & $0.65\pm 0.01$ & $-0.83$ & $-0.74$ & $0.65$ & $0.71$ & $0.72$ & $0.77$ & $\cdot \cdot \cdot $ \\ 
$\phi \rightarrow \eta ^{\prime }$ & $0.69\pm 0.02$ & $0.89$ & $0.83$ & $0.82$ & $-0.66$ & $0.71$ & $0.88$ & $-0.67$ \\ 
$\eta ^{\prime }\rightarrow \omega ^{0}$ & $0.40\pm 0.02$ & $0.62$ & $0.53$ & $0.50$ & $0.63$ & $0.58$ & $\cdot \cdot \cdot $ & 
$0.49$ \\ 
$\eta ^{\prime }\rightarrow \rho ^{0}$ & $1.23\pm 0.01$ & $1.97$ & $1.75$ & $1.68$ & $1.85$ & $1.69$ & $\cdot \cdot \cdot $ & 
$1.48$ \\ 
$D^{*+}\rightarrow D^{+}$ & $0.44\pm 0.05$ & $-0.54$ & $-0.45$ & $-0.40$ & $-0.35$ & $0.49$
& $0.44$ & $\cdot \cdot \cdot $ \\ 
$D^{*0}\rightarrow D^{0}$ & $\cdot \cdot \cdot $ & $2.25$
& $1.82$ & $1.68$ & $1.78$ & $1.64$ & $1.2$ & $\cdot \cdot \cdot $ \\ 
$D_{s}^{*+}\rightarrow D_{s}^{+}$ & $\cdot \cdot \cdot $
& $-0.22$ & $-0.26$ & $-0.23$ & $-0.13$ & $0.25$ & $0.26$ & $\cdot \cdot \cdot $ \\  
$J/\psi \rightarrow \eta _{c}$ & $0.65\pm 0.09$ & $0.78$ & $0.60$ & $0.59$ & $0.69$ & $%
\cdot \cdot \cdot $ & $\cdot \cdot \cdot $ & $0.59$ \\ 
$B^{*+}\rightarrow B^{+}$ & $\cdot \cdot \cdot $ & $1.80$
& $1.36$ & $1.31$ & $-1.37$ & $1.39$ & $1.63$ & $\cdot \cdot \cdot $ \\ 
$B^{*0}\rightarrow B^{0}$ & $\cdot \cdot \cdot $ & $%
-0.99 $ & $-0.76$ & $-0.74$ & $-0.78$ & $0.72$ & $0.92$ & $\cdot \cdot \cdot $ \\ 
$B_{s}^{*0}\rightarrow B_{s}^{0}$ & $\cdot \cdot \cdot $
& $-0.67$ & $-0.59$ & $-0.58$ & $-0.55$ & $0.47$ & $0.71$ & $\cdot \cdot \cdot $ \\ 
$B_{c}^{*+}\rightarrow B_{c}^{+}$ & $\cdot \cdot \cdot $
& $0.33$ & $0.24$ & $0.24$ & $\cdot \cdot \cdot $ & $0.33 $ & $0.31$ & $\cdot \cdot \cdot $ \\ 
$\Upsilon
\rightarrow \eta _{b}$ & $\cdot \cdot \cdot $ & $-0.12$ & $-0.11$ & $-0.11$
& $-0.13$ & $\cdot \cdot \cdot $ & $\cdot \cdot \cdot $
& $-0.09$ \\
\noalign{\smallskip}\hline\noalign{\smallskip}
\multicolumn{9}{l}{$^{a}$~Only absolute values $\left| \mu \right| $ are
presented.}
\end{tabular}
\end{table*}%

To proceed with the calculations we must specify the photon momenta $k.$ In
order to reduce possible uncertainties these momenta have been calculated
using experimental mass values of the mesons under consideration. The only
exception is the vector meson $B_{c}^{*}$ whose experimental mass is still
missing. In this case the theoretical estimate is necessary. For heavy
mesons the photon momentum $k$ approximately coincides with the
corresponding mass difference. The typical quark model estimate of the
hyperfine splitting for the $B_{c}$ meson is $68\pm 8~\mathrm{MeV,}$ with an
alternative method giving $84~\mathrm{MeV}$ \cite{KR14}. The lattice QCD
prediction is $M_{B_{c}^{*}}-M_{B_{c}}=54\pm 3~\mathrm{MeV}$, very close to
the hyperfine splittings of the heavy-light mesons $%
M_{B_{s}^{*}}-M_{B_{s}}=52\pm 3~\mathrm{MeV}$ and $M_{B^{*}}-M_{B}=50\pm 3~%
\mathrm{MeV}$ \cite{DDHH12}. This is the definite indication that the $\overline{b}c$ system
behaves in some ways more like a heavy-light than a heavy-heavy one \cite
{MDFHL12}. Nevertheless, we know that lattice QCD predictions, as a rule,
tend to underestimate the masses and hyperfine splittings of heavy mesons.
On the other hand, the well elaborated relativistic potential model 
\cite{EFG03}, which provides a good agreement with the data for the
hyperfine splittings of other heavy mesons, predicts the splitting $%
M_{B_{c}^{*}}-M_{B_{c}}$ a few $\mathrm{MeV}$ above the splitting 
$M_{\Upsilon}-M_{\eta _{b}}$ in bottonium. We follow this suggestion, take the
experimental value $M_{\Upsilon}-M_{\eta _{b}}=62.3\pm 3.2~\mathrm{MeV}$, add $2~\mathrm{MeV}$, and obtain
the estimate $M_{B_{c}^{*}}-M_{B_{c}}\approx 64~\mathrm{MeV}$ in rough
agreement with the modern quark model prediction $68\pm 8~\mathrm{MeV}$ \cite
{KR14}. The dependence of the transition moments on $k$ in the case of heavy
hadrons is relatively slow, therefore for the calculation of these
observables the exact value of $k$ is not very important. On the other hand,
the transition rates behave as $k^{3}$. Therefore, in the calculation of the
decay width of $B_{c}^{*+}$ meson the main source of uncertainty is the
ambiguity in the choice of the photon momentum.

\begin{table*}[tbp] \centering%
\caption{ M1 decay widths (in keV) of ground-state vector mesons.} 
\label{cal3.03}
\begin{tabular}{lccccccccc}
\hline\noalign{\smallskip}
Transition & Experiment & Our & Bag & CBM & QM & RQM & RPM & RPM & LFQM \\ 
& \cite{PDG14} &  & \cite{HDDT78} & \cite{SM86,MS88,SM89} & \cite{KX92} & 
\cite{J91,J96} & \cite{JPT02} & \cite{CK01} & \cite{CJ99,C07} \\ 
\noalign{\smallskip}\hline\noalign{\smallskip}
$\rho ^{+}\rightarrow \pi ^{+}$ & $67\pm 7$ & $66.7$ & $%
43.45$ & $124$ & $74.6$ & $76$ & $64.78$ & $53.3$ & $69$ \\ 
$\rho ^{0}\rightarrow \pi ^{0}$ & $90\pm 12$ & $66.7$ & $%
43.45$ & $124$ & $74.6$ & $76$ & $65.45$ & $53.7$ & $69$ \\ 
$\omega ^{0}\rightarrow \pi ^{0}$ & $703\pm 24$ & $616$
& $398.7$ & $1180$ & $716$ & $730$ & $613.3$ & $498$ & $667$ \\ 
$\omega ^{0}\rightarrow \eta $ & $3.9\pm 0.4$ & $7.96$ & 
$6.36$ & $2.3$ & $8.1$ & $8.7$ & $4.90$ & $5.28$ & $6.3\pm 0.3$ \\ 
$\rho ^{0}\rightarrow \eta $ & $45\pm 3$ & $55.7$ & $%
58.33$ & $23$ & $57.5$ & $59$ & $51.96$ & $51.6$ & $54$ \\ 
$K^{*+}\rightarrow K^{+}$ & $50\pm 5$ & $59.8$ & $7.71$
& $47$ & $82.3$ & $50$ & $58.10$ & $52$ & $71.4$ \\ 
$K^{*0}\rightarrow K^{0}$ & $116\pm 10$ & $118$ & $93.72$
& $98$ & $114$ & $117$ & $112.3$ & $125$ & $116.6$ \\ 
$\phi \rightarrow \pi ^{0}$ & $5.4\pm 0.3$ & $5.31$ & $0$
& $4.7$ & $5.8$ & $5.6$ & $4.89$ & $2.93$ & $2.5-8.7$ \\ 
$\phi \rightarrow \eta $ & $55.8\pm 1.4$ & $54.7$ & $%
43.72$ & $43$ & $31.3$ & $55.3$ & $68.23$ & $78$ & $47.6\pm 1.5$ \\ 
$\phi \rightarrow \eta ^{\prime }$ & $0.267\pm 0.011$ & $%
0.384$ & $2.39$ & $0.29$ & $0.34$ & $0.57$ & $0.30$ & $0.46$ & $0.34\pm 0.01$
\\ 
$\eta ^{\prime }\rightarrow \omega ^{0}$ & $5.4\pm 0.5$
& $8.72$ & $0$ & $6.0$ & $4.8$ & $4.8$ & $11.53$ & $13$ & $7.0\pm 0.4$ \\%
 
$\eta ^{\prime }\rightarrow \rho ^{0}$ & $57.6\pm 1.0$ & 
$108$ & $0$ & $53$ & $67.3$ & $67.5$ & $117.6$ & $117$ & $62$ \\ 
$D^{*+}\rightarrow D^{+}$ & $1.33\pm 0.33$ & $1.10$ & $%
0.82$ & $1.7$ & $1.42$ & $0.56$ & $1.63$ & $1.36$ & $0.90\pm 0.02$ \\ 
$D^{*0}\rightarrow D^{0}$ & $\cdot \cdot \cdot $ & $19.7$
& $22.57$ & $18.2$ & $21.7$ & $21.69$ & $19.48$ & $10.25$ & $20.0\pm 0.3$ \\%
 
$D_{s}^{*+}\rightarrow D_{s}^{+}$ & $\cdot \cdot \cdot $
& $0.40$ & $0.12$ & $0.10$ & $0.21$ & $\cdot \cdot \cdot $ & $0.44$ & $0.48$
& $0.18\pm 0.01$ \\ 
$J/\psi \rightarrow \eta _{c}$ & $1.58\pm 0.42$ & $1.31$
& $21.00$ & $2.0$ & $1.27$ & $\cdot \cdot \cdot $ & $\cdot \cdot \cdot $ & $%
\cdot \cdot \cdot $ & $1.69\pm 0.05$ \\ 
$B^{*+}\rightarrow B^{+}$ & $\cdot \cdot \cdot $ & $%
0.459 $ & $\cdot \cdot \cdot $ & $0.62$ & $\cdot \cdot \cdot $ & $0.429$ & $%
0.52$ & $0.71$ & $0.40\pm 0.03$ \\ 
$B^{*0}\rightarrow B^{0}$ & $\cdot \cdot \cdot $ & $%
0.146 $ & $\cdot \cdot \cdot $ & $0.28$ & $\cdot \cdot \cdot $ & $0.142$ & $%
0.14$ & $0.22$ & $0.13\pm 0.01$ \\ 
$B_{s}^{*0}\rightarrow B_{s}^{0}$ & $\cdot \cdot \cdot $
& $0.102$ & $\cdot \cdot \cdot $ & $0.10$ & $\cdot \cdot \cdot $ & $\cdot
\cdot \cdot $ & $0.06$ & $0.15$ & $0.068\pm 0.017$ \\ 
$B_{c}^{*+}\rightarrow B_{c}^{+}$ & $\cdot \cdot \cdot $ & $0.041$
& $\cdot \cdot \cdot $ & $\cdot \cdot \cdot $ & $\cdot \cdot \cdot $ & $%
\cdot \cdot \cdot $ & $0.03$ & $0.032$ & $\cdot \cdot \cdot $ \\ 
\noalign{\smallskip}\hline
\end{tabular}
\end{table*}%

\begin{table*}[tbp] \centering%
\caption{ M1 decay widths (in keV) of heavy vector mesons.} 
\label{cal3.04}
\begin{tabular}{lccccccccccc}
\hline\noalign{\smallskip}
Transition & Experiment & Our & HB & HB & QCDSR & LCSR & BSLT & RPM & $\chi $%
RQM & $\chi $EFT & PM \\ 
 & \cite{PDG14} &  & \cite{IDS82} & \cite{OH99} & \cite
{ZHY97} & \cite{ADIP96} & \cite{LNR00} & \cite{EFG02} & \cite{GR01} & \cite
{BEH03} & \cite{EGKLY80} \\ 
\noalign{\smallskip}\hline\noalign{\smallskip}
$D^{*+}\rightarrow D^{+}$ & $1.33\pm 0.33$ & $1.10$ & $%
0.9$ & $1.72$ & $0.23\pm 0.10$ & $1.50$ & $1.10$ & $1.04$ & $1.5$ & $1.63$ & 
$2.4$ \\ 
$D^{*0}\rightarrow D^{0}$ & $\cdot \cdot \cdot $ & $19.7$
& $20$ & $7.18$ & $12.9\pm 2.0$ & $14.40$ & $1.25$ & $11.5$ & $32\pm 1$ & $%
33.5$ & $35.2$ \\ 
$D_{s}^{*+}\rightarrow D_{s}^{+}$ & $\cdot \cdot \cdot $
& $0.40$ & $0.5$ & $\cdot \cdot \cdot $ & $0.13\pm 0.05$ & $\cdot \cdot
\cdot $ & $0.337$ & $0.19$ & $0.32\pm 0.01$ & $0.43$ & $0.32$ \\ 
$B^{*+}\rightarrow B^{+}$ & $\cdot \cdot \cdot $ & $0.46$
& $1.3$ & $0.272$ & $0.38\pm 0.06$ & $0.63$ & $0.0674$ & $0.19$ & $0.74\pm
0.09$ & $0.78$ & $1.7$ \\ 
$B^{*0}\rightarrow B^{0}$ & $\cdot \cdot \cdot $ & $0.15$
& $0.5$ & $0.064$ & $0.13\pm 0.03$ & $0.16$ & $0.0096$ & $0.070$ & $0.23\pm
0.03$ & $0.24$ & $0.5$ \\ 
$B_{s}^{*0}\rightarrow B_{s}^{0}$ & $\cdot \cdot \cdot $
& $0.10$ & $0.3$ & $0.051$ & $0.22\pm 0.04$ & $\cdot \cdot \cdot $ & $0.148$
& $0.054$ & $0.14\pm 0.02$ & $0.15$ & $0.2$ \\ 
\noalign{\smallskip}\hline
\end{tabular}
\end{table*}

\begin{table*}[tbp] \centering
\caption{Decay width (in keV) of M1 transition $J/\psi \rightarrow
\eta _{c}$.} 
\label{cal3.05}
\begin{tabular}{ccccccccccccc}
\hline\noalign{\smallskip}
Experiment & Our & PM & PM & PM & SRPM & RPM & BSLT & BSF & QCDSR & QCDSR & 
QCDSR & Latt \\ 
\cite{PDG14} &  & \cite{EGKLY80} & \cite{ZSG91} & \cite{BeA05} & \cite{BGS05}
& \cite{EFG03} & \cite{L03} & \cite{HS81} & \cite{S80} & \cite{A84} & \cite
{BR85} & \cite{DeA12a} \\
\noalign{\smallskip}\hline\noalign{\smallskip}
$1.58\pm 0.42$ & $1.31$ & $1.23$ & $1.8$ & $1.96$ & $2.4$ & $1.05$ & $1.25$
& $1.7$ & $2.7\pm 0.5$ & $2.1\pm 0.4$ & $2.6\pm 0.5$ & $2.49\pm 0.19$ \\ 
\noalign{\smallskip}\hline
\end{tabular}
\end{table*}

The predictions obtained using our extended bag model are listed in the
column of table~\ref{cal3.02} denoted as Our. It could be useful to
check what would happen in the long wavelength limit $k=0$. The
predictions obtained in such simplified version of the model with the same
other parameters ($C_{L}=1.43$, $C_{H}=0.87$, $\alpha _{V}=-4.1^{\circ }$,
and $\alpha _{P}=-45^{\circ }$) are listed in the column LWL of this table.
In the column denoted as NR we present
predictions of the simple nonrelativistic model, in which the quark
transition moments are replaced by the static magnetic moments. We take the
input values of the quark magnetic moments from ref.~\cite{FLNC81}. They are $\mu _{u}=1.86~\mu _{N}$, $\mu
_{d}=-0.93~\mu _{N}$, $\mu _{s}=-0.61~\mu _{N}$, $\mu _{c}=0.39~\mu _{N}$,
and $\mu _{b}=-0.06~\mu _{N}$. Note that
these values are adjusted to reproduce the magnetic moments of light
baryons. The vector mixing angle obtained from the fit to the experiment
now is $%
\alpha _{V}=-2.6^{\circ }$, and the pseudoscalar mixing angle $\alpha
_{P}=-45^{\circ }$ is left unchanged. Evidently these nonrelativistic results are not
acceptable for a serious comparison with experiment. Of course, they could
serve as a kind of reference point revealing some quark model problems,
but there is one useful exception. For the bottonium the nonrelativistic
description is undoubtedly a good approximation, and therefore we expect the
transition moment $\mu _{\Upsilon \rightarrow \eta _{b}}$ obtained in the nonrelativistic quark model to be of sufficiently high accuracy.

In table~\ref{cal3.02} we also compare our predictions with the results
obtained using other approaches, such as:

\begin{itemize}
\item  the semi-relativistic (relativized) potential model (SRPM) \cite{GI85};

\item  relativistic potential models (RPM) \cite{JPT02,CK01};

\item  the statistical model (SM) \cite{BDS89}.
\end{itemize}

The experimental values of M1 transition moments have been deduced from the
partial decay widths given in Particle Data Tables \cite{PDG14} with the
help of relations inverse to eqs.~(\ref{bag 020}) and (\ref{bag 021}).

From table~\ref{cal3.02} we see that the results obtained in our present
model are of similar quality as those obtained using other approaches. Moreover,
apart from several exceptions, our predictions are close to the
predictions obtained in the relativistic potential model \cite{JPT02}.

The comparison with experimental data shows that in the light meson sector
for the transitions without most problematic mesons $\eta $ and $\eta
^{\prime }$ the agreement with data is satisfactory. In more detail, for the
transitions $\rho ^{+}\rightarrow \pi ^{+}$ and $K^{*0}\rightarrow K^{0}$
the agreement width the data is excellent. For the decay $\omega
^{0}\rightarrow \pi ^{0}$ a possible deviation of the calculated transition
moment from the experimental value can be $5-7\%$, for the decay $%
K^{*+}\rightarrow K^{+}$ about $10\%$, and for the transition $\rho
^{0}\rightarrow \pi ^{0}$ the possible deviation from the experiment can be as
large as $18\%$. The latter uncertainty is the largest. Note that this is a
common problem of all quark model based approaches. The isospin symmetry
implies $\mu _{\rho ^{+}\rightarrow \pi ^{+}}=\mu _{\rho ^{0}\rightarrow
\pi ^{0}}$, however, we see that the experimental values of these quantities
do not coincide. This could mean that in this case the isospin symmetry
breaking should be taken into account, and, as a consequence, the $\rho
^{0}$-$\omega ^{0}$ mixing is possible. If we ignore this largest
uncertainty, then typical deviation from experimental data does not exceed $%
10\%.$ We think this is a reasonable measure of the accuracy of the method.
We can also conclude that the typical accuracy of the calculated magnetic
(transition) moments of the light quarks is about $5\%$.

Because we have used a rather crude description of pseudoscalars $\eta $ and $%
\eta ^{\prime }$, in these cases we do not expect a very good agreement with
experiment. Our predictions for the transition moments of these mesons are to
some extent similar to the predictions obtained in other relativistic
models. In all cases the agreement of the predictions with the experimental data
is far from being perfect. Moreover, the predictions strongly depend on the
pseudoscalar mixing angle $\alpha _{P}$. We have analyzed the dependence of
the ratios $\mu _{\phi \rightarrow \eta ^{\prime }}/\mu _{\phi \rightarrow
\eta }$, $\mu _{\omega ^{0}\rightarrow \eta }/\mu _{\eta ^{\prime
}\rightarrow \omega ^{0}}$, and $\mu _{\rho ^{0}\rightarrow \eta }/\mu
_{\eta ^{\prime }\rightarrow \rho ^{0}}$ on the angle $\alpha _{P}$. The
experimental values are $\mu _{\phi \rightarrow \eta ^{\prime }}/\mu _{\phi
\rightarrow \eta }=1.06\pm 0.03$, $\mu _{\omega ^{0}\rightarrow \eta }/\mu
_{\eta ^{\prime }\rightarrow \omega ^{0}}=1.05\pm 0.09$, $\mu _{\rho
^{0}\rightarrow \eta }/\mu _{\eta ^{\prime }\rightarrow \rho ^{0}}=1.21\pm
0.05$. The first two are compatible with the predictions obtained with $%
\alpha _{P}=-40^{\circ }$, while $\mu _{\rho ^{0}\rightarrow \eta }/\mu
_{\eta ^{\prime }\rightarrow \rho ^{0}}=1.21$ requires $\alpha _{P}=%
-51^{\circ }$. So we can not fit all three ratios simultaneously, and the
perfect mixing angle $\alpha _{P}=-45^{\circ }$ seems to be an
acceptable compromise.

In the heavy meson sector our predictions are in satisfactory agreement with the
available experimental data (transition moments $\mu _{D^{*+}\rightarrow
D^{+}}$ and $\mu _{J/\psi \rightarrow \eta _{c}}$). Moreover, we have one
additional reliable entry to compare with, \textit{i.e.}, the nonrelativistic value of the
transition moment $\mu _{\Upsilon
\rightarrow \eta _{b}}$. We see that our prediction $\mu _{\Upsilon
\rightarrow \eta _{b}}=-0.11~\mu _{N}$ is close enough to the
nonrelativistic one. Thus, adjusting one model parameter
(scale factor $C_{H}$) we have achieved the reliable predictions for three
transition moments, and this gives us some confidence that our further
predictions for the heavy mesons are more or less reliable, too. In the
heavy meson sector the serious source of uncertainty is the ambiguity in the
choice of the scale factor $C_{H}$. Most sensitive to this ambiguity are
small transition moments. The uncertainties are estimated to be $4\%$ for $%
\mu _{D^{*+}\rightarrow D^{+}}$, $9\%$ for $\mu _{D_{s}^{*+}\rightarrow
D_{s}^{+}}$, $5\%$ for $\mu _{J/\psi \rightarrow \eta _{c}}$, $7\%$ for $\mu
_{B_{c}^{*+}\rightarrow B_{c}^{+}}$, and $6\%$ for $\mu _{\Upsilon
\rightarrow \eta _{b}}$. The influence of this ambiguity on other transition
moments ($\mu _{D^{*0}\rightarrow D^{0}}$, $\mu _{B^{*+}\rightarrow B^{+}}$, 
$\mu _{B^{*0}\rightarrow B^{0}}$, and $\mu _{B_{s}^{*0}\rightarrow
B_{s}^{0}} $) does not exceed $1\%$, however, the real uncertainties for these
transitions could be larger (up to $6\%$) due to the error in the values of
transition moments of the light quarks.

\begin{table}[tbp] \centering%
\caption{Decay width (in eV) of M1 transition $B_{c}^{*+}\rightarrow
B_{c}^{+}$.} 
\label{cal3.06}
\begin{tabular}{ccccccccc}
\hline\noalign{\smallskip}
Our & PM & PM & BSLT & SRPM & RPM & RPM \\  
& \cite{GKLT95} & \cite{F99} & \cite{L03} & \cite{G04} & \cite{BD94,JPM98,JPT99,JPS01} & 
\cite{EFG03} \\
\noalign{\smallskip}\hline\noalign{\smallskip}
$41$ & $60$ & $59$ & $34$ & $80$ & $20-30$ & $33$ \\ 
\noalign{\smallskip}\hline
\end{tabular}
\end{table}

\begin{table}[tbp] \centering%
\caption{Decay width (in eV) of M1 transition $\Upsilon\rightarrow \eta
_{b}$.\label{cal3.07}} 
\label{cal3.07}
\begin{tabular}{cccccc}
\hline\noalign{\smallskip}
Our & PM & PM & BSLT & RPM & SIE \\  
& \cite{ZSG91} & \cite{BeA05} & \cite{L03} & \cite{EFG03} 
& \cite{ADMNS07} \\ 
\noalign{\smallskip}\hline\noalign{\smallskip}
$7.8$ & $4.0$ & $8.95$ & $7.7$ & $5.8$ & $7.9$ \\ 
\noalign{\smallskip}\hline
\end{tabular}
\end{table}%

From the comparison of our predictions for the transition moments of light
mesons with the results obtained in the long wavelength limit (LWL) and with
the results obtained using nonrelativistic quark model (NR) we see that it is
essential to take into account the $k$-dependence of these quantities in
order to achieve the satisfactory agreement with experimental data. In the
case of the mesons made of one charmed and one light quark the $k$-dependence of the transition moments is also important. For the
heaviest mesons ($J/\psi $, $B$, $B_{s}$, $B_{c}$, and $\Upsilon$) the results obtained in the long wavelength limit are similar to our
present predictions, and we conclude that for the M1 transitions of these
hadrons the long wavelength limit can be a reasonable approximation.

So far we were concentrated on the M1 transition moments, however, the
really measurable quantities are the transition rates. In order to have a
complete picture we have used the values of the transition moments presented
above as inputs to calculate the partial decay widths. They are listed in
table~\ref{cal3.03} and compared with the experimental data. We
also compare our predictions with the results obtained using various other
approaches. These are:

\begin{itemize}
\item  the bag model (Bag) \cite{HDDT78};

\item  the cloudy bag model (CBM) \cite{SM86,MS88,SM89}. The presented results for $D$ mesons correspond to the value of their scale factor $\lambda =0.7$;

\item  the simple quark model (QM) \cite{KX92};

\item  the relativistic quark model (RQM) \cite{J91,J96};

\item  relativistic potential models (RPM) \cite{JPT02,CK01};

\item  the light front quark model with linear confining potential (LFQM) \cite
{CJ99,C07}.
\end{itemize}

Given our special interest in the magnetic properties of heavy mesons, in
tables~\ref{cal3.04}--\ref{cal3.07} we continue the comparison of our
predictions with the results for the heavy meson sector obtained using the
following approaches:

\begin{itemize}
\item  bag models for heavy hadrons (HB) \cite{IDS82,OH99};

\item  usual QCD sum rules (QCDSR) \cite{ZHY97,S80,A84,BR85};

\item  light cone QCD sum rules (LCSR) \cite{ADIP96};

\item  potential models (PM) \cite{EGKLY80,ZSG91,GKLT95,F99};

\item  the formalism based on the
Blankenbecler-Sugar
equation (BSLT) \cite{L03,LNR00};

\item the Bethe-Salpeter formalism (BSF) \cite{HS81}.

\item  the emi-relativistic (relativized) potential model (SRPM) \cite
{G04,BGS05};

\item  relativistic potential models (RPM) \cite
{BD94,JPM98,JPT99,JPS01,EFG02,EFG03};

\item  the elativistic chiral quark model ($\chi $RQM) \cite{GR01};

\item  chiral effective field theory ($\chi $EFT) \cite{BEH03};

\item  lattice QCD calculations (Latt) \cite{DeA12a};

\item  the framework of the spectral integral equations (SIE) \cite{ADMNS07}  (the fit with retarded interactions).
\end{itemize}

The inspection of the predictions obtained using our approach and various
other methods shows that the overall agreement is not bad, nevertheless,
some mess-up is present especially for the transitions of heavy-light
mesons. We see the serious improvement in the predictions of the decay
widths as compared with the results obtained in the earlier versions of bag
model \cite{HDDT78,IDS82,SM86,MS88,SM89,OH99}. For the heavy mesons
similar results are obtained in our model and in ref.~\cite{IDS82} in the
charm sector, but in the bottom sector the predictions differ substantially. We stress
that our predictions in most cases agree with available experimental data
and are similar to the results obtained in the framework
of the relativistic potential model \cite{JPT02}. They are also more or less compatible with the predictions obtained in the light front quark
model \cite{CJ99,C07}. For the transitions of heavy mesons $J/\psi
\rightarrow \eta _{c}$, $B_{c}^{*+}\rightarrow B_{c}^{+}$, and 
$\Upsilon
\rightarrow \eta _{b}$ our predictions are very close to the results
obtained in ref.~\cite{L03} in the formalism based on the
Blankenbecler-Sugar equation. The prediction very close to our result has been also obtained for the $\Upsilon
\rightarrow \eta _{b}$ transition in the framework of the spectral integral equations (SIE) \cite{ADMNS07}. We note also some deviations from RPM results
obtained in refs.~\cite{EFG02,EFG03}: for the transitions $D^{*+}\rightarrow
D^{+}$, $J/\psi \rightarrow \eta _{c}$, $B_{c}^{*+}\rightarrow
B_{c}^{+}$, and $\Upsilon
\rightarrow \eta _{b}$ our predictions are large by about $30\%$, and for
other decays (see table~\ref{cal3.04}) they are approximately twice as
large.

\section{Magnetic moments}
\label{sec_rad}

As we have seen, the method developed above is capable to provide reasonable
predictions for M1 transition moments and corresponding partial decay rates
of ground-state vector mesons. Thus we can expect that the predictions for
usual magnetic moments obtained using the same method also should be
reliable. Below we list detailed expressions for these magnetic moments
(they are extremely simple and are presented only for convenience) 
\begin{equation}
\mu _{\rho ^{+}}=\widetilde{\mu }_{l}\,,  \label{rad 01}
\end{equation}
\begin{equation}
\mu _{K^{*+}}=\frac{2}{3}\widetilde{\mu }_{l}+\frac{1}{3}\widetilde{\mu }%
_{s}\,,  \label{rad 03}
\end{equation}
\begin{equation}
\mu _{K^{*0}}=-\frac{1}{3}\widetilde{\mu }_{l}+\frac{1}{3}\widetilde{\mu }%
_{s}\,,  \label{rad 04}
\end{equation}
\begin{equation}
\mu _{D^{*+}}=\frac{1}{3}\widetilde{\mu }_{l}+\frac{2}{3}\widetilde{\mu }%
_{c}\,,  \label{rad 05}
\end{equation}
\begin{equation}
\mu _{D^{*0}}=-\frac{2}{3}\widetilde{\mu }_{l}+\frac{2}{3}\widetilde{\mu }%
_{c}\,,  \label{rad 06}
\end{equation}
\begin{equation}
\mu _{D_{s}^{*+}}=\frac{2}{3}\widetilde{\mu }_{c}+\frac{1}{3}\widetilde{\mu }%
_{s}\,,  \label{rad 07}
\end{equation}
\begin{equation}
\mu _{B^{*+}}=\frac{2}{3}\widetilde{\mu }_{l}+\frac{1}{3}\widetilde{\mu }%
_{b}\,,  \label{rad 08}
\end{equation}
\begin{equation}
\mu _{B^{*0}}=-\frac{1}{3}\widetilde{\mu }_{l}+\frac{1}{3}\widetilde{\mu }%
_{b}\,,  \label{rad 09}
\end{equation}
\begin{equation}
\mu _{B_{s}^{*0}}=-\frac{1}{3}\widetilde{\mu }_{s}+\frac{1}{3}\widetilde{\mu 
}_{b}\,,  \label{rad 10}
\end{equation}
\begin{equation}
\mu _{B_{c}^{*+}}=\frac{2}{3}\widetilde{\mu }_{c}+\frac{1}{3}\widetilde{\mu }%
_{b}\,.  \label{rad 11}
\end{equation}
Note that $\mu _{\rho ^{0}}=\mu _{\omega ^{0}}=\mu _{J/\psi }=\mu _{%
\Upsilon}=0$.

\begin{table*}[tbp] \centering 
\caption{Magnetic moments of light mesons in nuclear
magnetons (values in natural magnetons are given in parenthesis).}
\label{rad4 01}
\begin{tabular}{lccccccccc}
\hline\noalign{\smallskip}
Particle & Our & NR & RH & DSE & DSE & Latt & Latt & LCSR & NJL \\ 
&  &  & \cite{BS13} & \cite{BM08} & \cite{HP99} & \cite{LMW08} & \cite
{HKLLWZ07} & \cite{AOS09} & \cite{LCD15} \\ 
\noalign{\smallskip}\hline\noalign{\smallskip}
$\rho ^{+}$ & $2.50$ & $2.79$ & $2.37$ & $2.43$ & $3.28$ & $3.25\pm 0.03$ & $%
2.3$ & $2.9\pm 0.5$ & $2.54$ \\
`` & $(2.06)$ & $(2.31)$ & $(1.96)$ & $(2.01)$ & $(2.69)$ & $(2.39\pm 0.01)$
& $(2.2)$ & $(2.4\pm 0.4)$ & $(2.09)$ \\ 
\noalign{\smallskip}\hline\noalign{\smallskip}
$K^{*+}$ & $2.21$ & $2.47$ & $2.19$ & $2.34$ & $2.49$ & $2.81\pm 0.01$ & $2.1
$ & $2.1\pm 0.4$ & $2.26$ \\
`` & $(2.10)$ & $(2.35)$ & $(2.09)$ & $(2.23)$ & $(2.37)$ & $(2.38\pm 0.01)$
& $(2.0)$ & $(2.0\pm 0.4)$ & $(2.14)$ \\ 
\noalign{\smallskip}\hline\noalign{\smallskip}
$K^{*0}$ & $-0.216$ & $-0.32$ & $-0.183$ & $-0.27$ & $-0.42$ & $\cdot \cdot
\cdot $ & $-0.07$ & $0.29\pm 0.04$ & $\cdot \cdot \cdot $ \\
`` & $(-0.206)$ & $(-0.31)$ & $(-0.175)$ & $(-0.26)$ & $(-0.40)$ & $\cdot
\cdot \cdot $ & $(-0.07)$ & $(0.28\pm 0.04)$ & $\cdot \cdot \cdot $ \\ 
\noalign{\smallskip}\hline
\end{tabular}
\end{table*}

\begin{table*}[tbp] \centering 
\caption{Magnetic moment of $\rho ^{+}$ meson in nuclear
magnetons (values in natural magnetons are given in parenthesis).}
\label{rad4 02}
\begin{tabular}{lccccccccccc}
\hline\noalign{\smallskip}
Our & Experiment & Latt & EFT & QCDSR & DSE & LFQM & LFQM & LFQM & LFQM & LFQM & 
LFQM \\ 
& \cite{SG14} & \cite{OKLMM15} & \cite{DEGM14} & \cite{S03} & \cite{IKR99} & 
\cite{CGNSS95} & \cite{BCJ02} & \cite{CJ04} & \cite{dF97} & \cite{J03} & 
\cite{MS02} \\ 
\noalign{\smallskip}\hline\noalign{\smallskip}
$2.50$ & $2.54\pm 0.61$ & $2.61\pm 0.10$ & $2.71$ & $2.4\pm 0.4$ & $2.95$ & $%
2.75$ & $2.55$ & $2.34$ & $2.61$ & $2.24$ & $2.86$ \\ 
$(2.05)$ & $(2.1\pm 0.5)$ & $(2.21\pm 0.08)$ & $(2.24)$ & $(2.0\pm 0.3)$ & $%
(2.44)$ & $(2.26)$ & $(2.1)$ & $(1.92)$ & $(2.14)$ & $(1.83)$ & $(2.35)$ \\ 
\noalign{\smallskip}\hline
\end{tabular}
\end{table*} 

\begin{table}[tbp] \centering 
\caption{Magnetic moments (in nuclear
magnetons) of heavy mesons and ratios of these magnetic moments to that of the proton.}
\label{rad4 03}
\begin{tabular}{lrccrrrr}
\hline\noalign{\smallskip}
Particle & Our & BSLT & NJL & NR & Bag & $\mu _{i}/\mu _{P}$ & $\mu _{i}/\mu
_{P}$ \\ 
&  & \cite{L03} & \cite{LCD15} &  & \cite{BS80} & Our & \cite{BS80} \\ 
\noalign{\smallskip}\hline\noalign{\smallskip}
$D^{*+}$ & $1.06$ & $\cdot \cdot \cdot $ & $1.16$ & 
$1.32$ & $1.17$ & $0.38$ & $0.42$ \\
$D^{*0}$ & $-1.21$ & $\cdot \cdot \cdot $ & $\cdot \cdot
\cdot $ & $-1.47$ & $-0.89$ & 
$-0.43$ & $-0.32$ \\
$D_{s}^{*+}$ & $0.87$ & $\cdot \cdot \cdot $ & $0.98$ & 
$1.00$ & $1.03$ & $0.31$ & $0.37$ \\
$B^{*+}$ & $1.47$ & $\cdot \cdot \cdot $ & $1.47$ & 
$1.92$ & $1.54$ & $0.53$ & $0.55$ \\
$B^{*0}$ & $-0.65$ & $\cdot \cdot \cdot $ & $\cdot \cdot
\cdot $ & $-0.87$ & $-0.64$ & $-0.23$ & $-0.23$ \\
$B_{s}^{*0}$ & $-0.48$ & $\cdot \cdot \cdot $ & $\cdot
\cdot \cdot $ & $-0.55$ & $-0.47$ & 
$-0.17$ & $-0.17$ \\
$B_{c}^{*+}$ & $0.35$ & $0.426$ & $\cdot \cdot \cdot $ & 
$0.45$ & $0.56$ & $0.13$ & $0.20$ \\ 
\noalign{\smallskip}\hline
\end{tabular}
\end{table} 

The results of our calculations for the light mesons are presented in 
tables~\ref{rad4 01} and \ref{rad4 02} together with the estimates obtained
in nonrelativistic quark model (NR). These predictions are compared with the
estimates obtained in other approaches. These are:

\begin{itemize}
\item  the relativistic Hamiltonian model (RH) \cite{BS13};

\item  various models based on the Dyson-Schwinger equation (DSE) \cite
{BM08,IKR99,HP99};

\item  lattice QCD calculations (Latt) \cite{HKLLWZ07,LMW08,OKLMM15};

\item  approaches based on standard QCD sum rules (QCDSR) \cite{S03} and its
light cone modification (LCSR) \cite{AOS09};

\item  the extended Nambu-Jona-Lasinio model with heavy quarks (NJL) \cite{LCD15};

\item  the effective field theory (EFT) \cite{DEGM14};

\item  light front quark models (LFQM) \cite{BCJ02,CJ04,CGNSS95,J03,dF97,MS02}.
\end{itemize}

Also we include in the comparison (column denoted as Experiment in table~\ref
{rad4 02}) the attempt to extract the magnetic dipole moment of the $\rho ^{+}$
meson using preliminary experimental data from the BaBar Collaboration. 

In many papers the values of magnetic moments are presented in
natural particle's magnetons $e/(2M_{i})$, where $M_{i}$ represents the mass
of corresponding meson. We have converted them to the values expressed in
nuclear magnetons by multiplying with the factor $M_{P}/M_{i}$. The
exception is lattice QCD predictions \cite{HKLLWZ07,LMW08,OKLMM15}, where they
have used a specific extrapolation procedure.

From tables~\ref{rad4 01} and \ref{rad4 02} we see that our predictions
are somewhere between the predictions obtained within the relativistic
Hamiltonian formalism (RH) and results obtained in the simple nonrelativistic
quark model (NR), but closer to the RH predictions. They are also more or less compatible with the theoretical results obtained in ref.~\cite{BM08} using the approach based on the Dyson-Schwinger equation (DSE) and with quenched lattice predictions (ref.~\cite{HKLLWZ07}). Within the error bars our predictions for the $\rho ^{+}$
and $K^{*+}$ mesons also agree with the results obtained using light cone
QCD sum rules (LCSR) and are close to the predictions obtained in the framework of the extended Nambu-Jona-Lasinio model (NJL). On the other hand, predictions obtained using another
version of Dyson-Schwinger formalism (ref.~\cite{HP99}) and lattice
predictions obtained in ref.~\cite{LMW08} differ substantially from ours.

For the $\rho ^{+}$ meson our prediction is slightly above the effective
field theory tree level value $2.42~\mu _{N}$ ($2.0$ in natural
magnetons) and agrees within the error bars with the estimate extracted from
experimental data (ref.~\cite{SG14}). The recent full QCD prediction \cite
{OKLMM15} is also in good agreement with our result. Regarding the light
front quark models (LFQM), the predictions obtained for the $\rho ^{+}$  using various variants of LFQM cover rather large range from $2.24~\mu _{N}$ up to $2.86~\mu _{N}$. Our prediction is closest to
the results obtained in the covariant versions of LFQM~\cite{BCJ02,dF97}.

We finish our investigation with the predictions for the magnetic moments of
ground-state vector mesons containing heavy quarks. The results are listed
in table~\ref{rad4 03}. They are compared with the results obtained in
the nonrelativistic quark model (NR), the predictions obtained in the framework of extended Nambu-Jona-Lasinio model (NJL), and with the prediction for $B_{c}^{*+}$
meson obtained using the formalism based on the covariant Blankenbecler-Sugar equation
(BSLT) \cite{L03}. There also exists the old bag model prediction for the
ratios of these magnetic moments to the magnetic moment of the proton \cite
{BS80}. For comparison we have presented these ratios together with ours in
the two last columns of table~\ref{rad4 03}. Magnetic moments deduced
from these ratios by multiplying them with the magnetic moment of proton $\mu
_{P} $ are given in the column named as Bag.

We see that the values of the magnetic moments predicted using our extended
version of the bag model are appreciably smaller than 
nonrelativistic results. Our prediction for the $B_{c}^{*+}$ meson is also
smaller than the
corresponding prediction obtained using BSLT formalism, the latter being similar to the nonrelativistic one. For $D^{*+}$ and $D_{s}^{*+}$ mesons the values obtained in the NJL model are between the NR predictions and ours. For $B^{*+}$ our and NJL predictions coincide. We have found with some surprise that our
predictions for $D^{*+}$, $B^{*+}$, $B^{*0}$, and $B_{s}^{*0}$ mesons are similar to
the rescaled old bag model results. However, for other mesons ($D^{*0}$%
, $D_{s}^{*+}$, and $B_{c}^{*+}$) the difference is evident. 

The uncertainties for the magnetic moments are estimated to be of the same
order as for transition moments. In the light meson sector the reasonable
estimate of possible error could be about $5\%$ for $\mu _{\rho ^{+}}$, and
up to $10\%$ for $\mu _{K^{*+}}$ and $\mu _{K^{*0}}$. In the heavy meson
sector the largest uncertainty ($\approx 6\%$) is expected for the $\mu
_{B_{c}^{*+}}$. For all other magnetic moments of heavy mesons the possible
uncertainty is expected to be smaller than $5\%$.

\section{Discussion and summary}
\label{sec_sum}

We have developed a method to treat the magnetic observables (i.e., magnetic
moments, M1 transition moments, and partial M1 decay widths) of ground-state
vector mesons. The method is based on slightly modified bag model \cite{BS04}. The main
difference from our earlier approach \cite{BS04,BS13b} is the recipe how to
take into account the CMM corrections of the magnetic observables. In our
previous approach following the usual procedure these corrections were applied
to the magnetic observable as a whole. The present investigation has shown
that a more reliable approach is to apply these corrections at the quark level.

We have used this extended bag model to calculate magnetic moments and
partial M1 decay widths of all ground-state vector mesons. To our knowledge,
our current predictions for the magnetic moments of neutral mesons $D^{*0}$, 
$B^{*0}$, and $B_{s}^{*0}$ are the first
reliable theoretical estimates of these properties.

In order to test the method we have compared our predictions for M1
transition moments and partial decay widths with the experimental data and
with the results obtained using other approaches. We have found a satisfactory
agreement with experiment and, to some extent, with other theoretical
predictions. Nevertheless, some aspects concerning the heavy meson sector
are not completely clear. Theoretical predictions obtained in various approaches
are somewhat dispersed, and so far only two M1 decay widths of heavy mesons
have been measured experimentally. The accuracy of data also is not very
high. It is not utterly clear if with such a small amount of data the
reliability of any model in heavy meson sector can be tested with
sufficiently high accuracy. At present the agreement of our predictions with
available data is good. In addition, our prediction for the transition moment 
$\mu _{\Upsilon \rightarrow \eta _{b}}$ is similar to the more or less realistic result
obtained in the nonrelativistic quark model. As concerns the magnetic
moments, our prediction for the $\rho ^{+}$ meson agrees within the error
bars with the estimate extracted from experimental data (ref.~\cite{SG14})
and is in good agreement with the recent full QCD prediction \cite{OKLMM15}.
Agreement with the predictions obtained using other approaches for the light
mesons is also satisfactory. We see that our method is capable to provide
the reasonable predictions for various magnetic properties, such as M1
transition moments (together with the partial decay widths) of light and
heavy mesons and magnetic moments of light mesons. Encouraged by this
success we expect our predictions for the magnetic moments of heavy mesons
to be also trustworthy.

\section*{Acknowledgement}

The author is indebted to A. Deltuva for the support and valuable advices.

\end{document}